\pdfoutput=1
\documentclass[12pt]{article}
\usepackage{graphicx}
\usepackage{amsmath}
\usepackage{amssymb}
\usepackage{booktabs}

\newcommand{\bea}{\begin{eqnarray}}
\newcommand{\eea}{\end{eqnarray}}
\newcommand{\be}{\begin{equation}}
\def\bel#1{\begin{equation} \label{#1}}
\newcommand{\ee}{\end{equation}}
\newcommand{\ba}{\begin{align}}
\newcommand{\ea}{\end{align}}
\newcommand{\comments}[1]{}

\begin{document}

\begin{titlepage}
\begin{flushright}
\parbox[t]{1.4in}{DAMTP-2016-72}
\end{flushright}

\begin{center}

\vspace*{ 1.2cm}

{\Large \bf Is String Phenomenology an Oxymoron?}

\vskip 1.2cm

\renewcommand{\thefootnote}{}
\begin{center}
Fernando Quevedo$^{1,2}$
 \end{center}
\vskip .2cm
\renewcommand{\thefootnote}{\arabic{footnote}}

{\it \small $^1$ ICTP, Strada Costiera 11, Trieste 34014, Italy.\\
$^2$ DAMTP, University of Cambridge, Wilberforce Road, Cambridge, CB3 0WA, UK.\\[0.5cm]}



\end{center}

\vskip 0.8cm

\begin{center} {\bf ABSTRACT } \end{center}
A brief discussion is presented assessing  the achievements and challenges of string phenomenology: the subfield dedicated to study the potential for string theory to make contact with particle physics and cosmology. Building from the well understood case of the standard model as a very particular example within quantum field theory we highlight the very few generic observable implications of string theory, most of them inaccessible to low-energy experiments, and indicate the need to extract  
concrete scenarios and classes of models that could eventually be contrasted with searches in collider physics and other particle experiments as well as in cosmological observations. The impact that this subfield has had in mathematics and in a better understanding of string theory is emphasised as spin-offs of string phenomenology.   Moduli fields, measuring the size and shape of extra dimensions, are highlighted as generic low-energy remnants of string theory that can play a key role for supersymmetry breaking as well as for inflationary and post-inflationary early universe cosmology. It is argued that the answer to the question in the title should be, as usual, {\it No}. Future challenges for this field are briefly mentioned.
This essay is a contribution to the conference: ``Why Trust a Theory?'',  Munich, December 2015.
\end{titlepage}

\tableofcontents
\section{Introduction}\label{ra_sec1}

Although the scope of this conference: ``Why trust a theory?'', was much broader and touched general aspects of the philosophy of science, the main motivation that triggered this meeting was the debate about string theory and its difficulty to be tested experimentally. Over a period of more than 30 years a subfield of string theory has developed that aims to actually address potential observational issues of string theory. This subfield  is known as string phenomenology.
The aim of string phenomenology is well defined and very ambitious: to uncover  string theory scenarios that satisfy all particle physics and cosmological observations and hopefully  lead to measurable predictions in the short, medium or long term. For a comprehensive treatment of the field with a very complete set of references   see \cite{iu}.

It is important to emphasise that most string theorists do not work directly on string phenomenology since string theory, being a theory under development, has many other open questions that range from conceptual to technical and computational challenges. We may distinguish string phenomenology from what could be called ``string noumenology''  (as introduced in \cite{noumenom}\, and appropriate for this conference) using the difference that philosophers like Kant make between {\it the noumenom} (the thing in itself)  and  {\it the phenomenom} (the thing as it manifests). Given the fact that string theory has been questioned precisely for its lack of concrete predictions that can be tested experimentally, it is natural to wonder about the term string phenomenology and this is what  motivates the title of this essay. We will argue that the answer to the question in the title is {\it No}. But before giving the arguments  we will recall some basic facts.

During the past 25 years we have witnessed at least four major discoveries that have had a big influence in understanding our universe:
\begin{enumerate}
\item{} The discovery of the density perturbations of the cosmic microwave background in 1992 by COBE which has been followed by the impressive precision of the  WMAP and PLANCK satellites
manifested by  the by-now famous plot of the power spectrum  of the primordial density perturbations giving rise to precise values of physical observables such as  the spectral index $n_s = 0.968 \pm 0.006$.

\item{} The surprising discovery in 1997 that our universe is not  only expanding but also accelerating indicating the existence of an unknown subject, dark energy, that overcomes the well known gravitational attraction and forces the universe to expand faster. The simplest manifestation of dark energy is a positive vacuum energy or cosmological constant (CC)  which is extremely small $\Lambda\sim 10^{-120} M_{Planck}^4$ \footnote{Here, $M_{Planck}=\sqrt{\hbar c/G}\sim 10^{19}$ GeV is the natural scale for relativistic ($c$) quantum ($\hbar$) gravity ($G$) known as the Planck scale.
}. The dark energy   equation of state is currently given by
$w=p/\rho = -1.006 \pm 0.045$. A positive cosmological constant giving rise to de Sitter space would imply $w=-1$. 
\item{} The  discovery of the Higgs boson in 2012, closing the particle content of the standard model (SM) of particle physics . The Higgs is responsible for the symmetry breaking mechanism that drives the electroweak unification and provides the mass scale for the standard model $m_{Higgs}=125$ GeV ($\sim 10^{-16} M_{Planck}$).

\item{} The spectacular discovery of gravitational waves by the  LIGO interferometers earlier  this year produced by merging black holes $1.2$ billion years ago. 
 
\end{enumerate}

These few major discoveries may be complemented by a few others probably less spectacular such as the discovery of the top quark, the accumulated evidence  for  non-vanishing neutrino masses, the precision tests of the standard model by LEP giving rise to the limit of three light neutrinos and providing indirect evidence for an intriguing apparent unification of the three gauge couplings of the standard model at a scale of order $M_{GUT}\sim10^{-2} M_{Planck}$ if the standard model (SM) is extended by a supersymmetric version. Also, the  indirect accumulation of experimental evidence for the existence of dark matter (although its nature is not yet uncovered). 
It should  be emphasised that these are success stories due to a community effort from theory, computational physics and experiment. There have also been, so far unsuccessful, attempts like the search for proton decay,  cosmic strings  (both by direct search through gravitational lensing and by their potential contribution to the CMB power spectrum), dark matter particles, axions, etc.

The discoveries mentioned above illustrate several important lessons regarding scientific methodology. The acceleration of the universe was hardly foreseen. Physicists had been trying for decades to address what now seems to be the wrong question, why the energy of the vacuum vanishes, with the remarkable exception of Weinberg's `prediction' in 1987 of the approximately observed experimental value through an anthropic argument \cite{WeinbergAnthropic}. Explicit calculations of the density perturbations were done only less than 15  years before their discovery whereas the potential existence of  the Higgs particle was predicted 50 years before and gravitational waves 100 years before being discovered. On this it is worth emphasising that Einstein predicted gravitational waves having no idea on how to detect them. Furthermore, only one year later he theoretically discovered  `stimulated emission'  which is the physical basis behind the lasers. But it would take 40 years for the first laser to be built and Einstein could not have dreamed that these devices, built following his idea and which he did not live to see, were going to be used to discover the gravitational waves he had predicted. This is a good example to keep in mind when discussing potential experimental tests of string theory. We have to be patient and at present we may not even imagine the 
technology to be used to eventually test it.

On the theoretical side, however, we cannot actually identify a similar list of greatest discoveries in the past 25 years. The nature and level of maturity of the field itself makes it difficult to state with certainty which theoretical developments will have the biggest impact in the relatively short term. Maybe only the gauge/gravity and possibly other string dualities are expected to survive a future assessment of important recent developments in high energy physics and it could be that discoveries in other fields, such as condensed matter or quantum information, or even other attempts to address the quantum gravity problem, may turn out to be more relevant. 

Even if we go one decade further, the 1980's,  we may only identify cosmological inflation (including the calculation of density perturbations)  and the emergence of string theory as a fundamental theory of nature as probably the main candidates to survive as great discoveries. Just before that, 
developments in gravity such as the Bekenstein/Hawking work in black hole thermodynamics are good contenders.  But we have to go to the  early 1970's to identify theoretical discoveries (such as the renormalisation of gauge theories and asymptotic freedom) that we are sure will remain in the long term. 

The points above illustrate that  big discoveries in theory are very rare and hard to identify in the short term. But these discoveries come from an accumulated effort by the whole community with small steps (and some failed attempts) building-up before  a major breakthrough. In the end, experimental evidence is the final judge and the timing depends on many factors, most of them beyond the control of the theorists. Therefore other criteria, such as mathematical consistency and well posed questions play an important role in guiding the research. These have also been the working premises in string theory so far.

\section{Basic Theories}

Let us start with an overview of the basic principles that are considered as the pillars of our current fundamental understanding of nature. The basic theories in physics so far are special relativity and quantum mechanics.
In 1939 Wigner took precisely these two theories and studied the representations of the Poincar\'e group (basis of special relativity) to identify the fundamental quantum states. They  are classified by the quantum numbers specified first by the eigenvalues of the Casimir operators $C_1=P^\mu P_\mu$ and $C_2=W^\mu W_\mu$ where $P^\mu,$ $\, \mu= 0,1,2,3$ is the momentum operator generator of space-time translations and $W^\mu=\epsilon^{\mu\nu\rho\sigma}P_\nu M_{\rho\sigma}$ where $M_{\rho\sigma}$ are the generators of rotations and Lorentz boosts and $W^\mu$ is the Pauli-Ljubanski vector. Besides the eigenvalues of $C_1$ and $C_2$ other labels identifying quantum states are given by the eigenvalues of generators that commute with $C_1, C_2$ and among themselves,  namely the momentum operators and other operators that commute with them. This naturally leaves three classes of states, depending on the sign of $C_1 $. 

If $C_1=m^2>0$ we can choose a frame for which the eigenvalues of the momenta are in the rest frame $p^\mu=(m,0,0,0)$ and the rest of the labels are determined by representations of the Little or Stability group $H$ defined as the subgroup of the Poincar\'e group that leaves invariant the corresponding momenta. In this case it is easy to identify  $H=O(3)$ as the group of rotations in three spatial dimensions. The finite dimensional unitary representations of $O(3)$ are well known to be determined by the eigenvalues of spin $J^2\propto C_2$  with integer or half-integer values and its third component $s=-J, -J+1,\cdots , J-1, J$. Therefore the quantum states are the one-particle states $|m, J; p^\mu, s\rangle $ essentially defining a massive particle. 

\be
{\rm Massive\, \, particle \, \, states} \qquad\qquad |m, J; p^\mu, s\rangle,\qquad s=-J, -J+1,\cdots, J-1, J
\ee

Further labels can be added corresponding to conserved quantities associated to internal  symmetries, such as electric charge, colour, etc. 

A similar analysis could be done for $C_1<0$. In this case the  states are tachyonic  corresponding to particles moving faster than the speed of light or $m^2<0$. These particles are not realised in nature but when appearing in a physical theory can usually be understood as perturbations of a field around a maximum (unstable) rather than a minimum vacuum state. 

Finally there are the massless states $C_1=0$ for which the momenta can be chosen as $p^\mu = (E,0,0,E)$. In this case the Little Group is the full Euclidean group in two dimensions which has unitary {\it infinite} dimensional representations. However we do not observe such a big degeneracy of massless particles. Imposing finite dimensional representations the Little group reduces to $O(2)$ which then adds one label to the representations corresponding to helicity $\lambda$. Topological conditions restrict helicity to be $\lambda =0,\pm \tfrac{1}{2},\pm 1, \pm \tfrac{3}{2}, \pm 2, \cdots $. In this case the eigenvalues of both $C_1$ and $C_2$ vanish and the states are simply labelled by $|p^\mu, \lambda \rangle $.

\be
{\rm Massless\, \,  particle\,  states} \qquad\qquad  |p^\mu, \lambda \rangle \qquad \lambda =0,\pm \tfrac{1}{2},\pm 1, \pm \tfrac{3}{2}, \pm 2, \cdots
\ee

The theory that describes interactions for massive and massless particles up to helicity $|\lambda| \leq  1$ is Quantum Field Theory (QFT). Gauge field theories in their different phases provide the consistent framework to describe interactions of particles of helicities $\lambda = \pm 1$. For higher helicities we need effective field theories (EFT). Effective field theories are well posed quantum field theories that are valid up to a cut-off scale $M$. General relativity describing $\lambda =\pm 2$ is an example. Supergravity, needed to also describe helicities $\lambda = \pm 3/2$ is another example.   In the 1960's Weinberg proved that particles of helicities $|\lambda| \geq 1$ need to couple to conserved currents. Particles of helicity $\lambda =\pm 1$ couple to gauge currents, those with $\lambda =\pm 2$ couple to the  stress energy tensor, $\lambda=\pm 3/2$ to the supersymmetry current. There are no further conserved quantities (from Coleman Mandula theorem and extensions) and therefore massless particles of higher helicities are not expected to exist, or at least do not mediate long range interactions. 

These ``soft theorems'' of Weinberg are very powerful and only rely on the basic understanding of special relativity and quantum mechanics. In particular, they  set the General Relativity description of gravity  at the same level of other interactions as just the  relativistic theory describing massless particles of helicity $\lambda=\pm 2$. This is the formulation of gravity that fits with string theory. For a detailed description of this perspective see for instance Weinberg's QFT textbook, in particular chapter 13 of volume 1 \cite{weinberg} (see also the last 2012   Salam Lecture of N. Arkani-Hamed: https://www.ictp.it/about-ictp/salam-lecture-series.aspx).

.

\section{General predictions of Quantum Field Theories}

By now there is no doubt that QFT is the right framework to address interactions of particles of spin one and below. Most of the work on QFT has been based on concrete models such as QED, QCD, etc.
But there are a handful of general predictions that can be extracted from all relativistic quantum field theories.

\begin{itemize}
\item{} {\it Identical particles}. All particles of the same type are indistinguishable from each other. Every electron is identical to any other electron and if one replaces one by the other in a material object no change should be noticed. This is simply because the electrons are all excitations of the same field (electron field), same for photons (electromagnetic field), etc. This is essentially a tautology since a particle is defined by its labels (mass, spin, etc.).

\item{} {\it Antiparticles}. Every field that creates a particle,  annihilates another particle with the same mass but opposite electric charge or other quantum numbers in the sense that they can annihilate each other in radiation. Some particles like the photon are their own antiparticles. This general implication of field theories has been verified experimentally since the discovery of the positron in 1939.

\item{} {\it Spin-statistics}. Particles of integer spin (bosons) behave very differently from those of half-integer spin (fermions) that are constrained by the Pauli exclusion principle. Bosons satisfy the Bose-Einstein statistics allowing them to be in the same state as in a laser beam whereas fermions are subject to the Fermi-Dirac statistics and the exclusion principle does not allow two of them in the same physical state, explaining the rich atomic structure of matter.

\item{} {\it The CPT theorem}. Physics is the same under a combination of time translations (T), space inversions (P) and charge conjugation (C).

\item{} {\it Decoupling and RG}. Physical processes are understood by different scales,  following the Wilsonian approach to field theory in which integrating out high scales describes an effective field theory for the lightest states. Couplings (such as gauge couplings and masses of particles) change with energy according to the renormalisation group (RG) equations.

\end{itemize}

\begin{figure}
\centerline{\includegraphics[height=8cm]{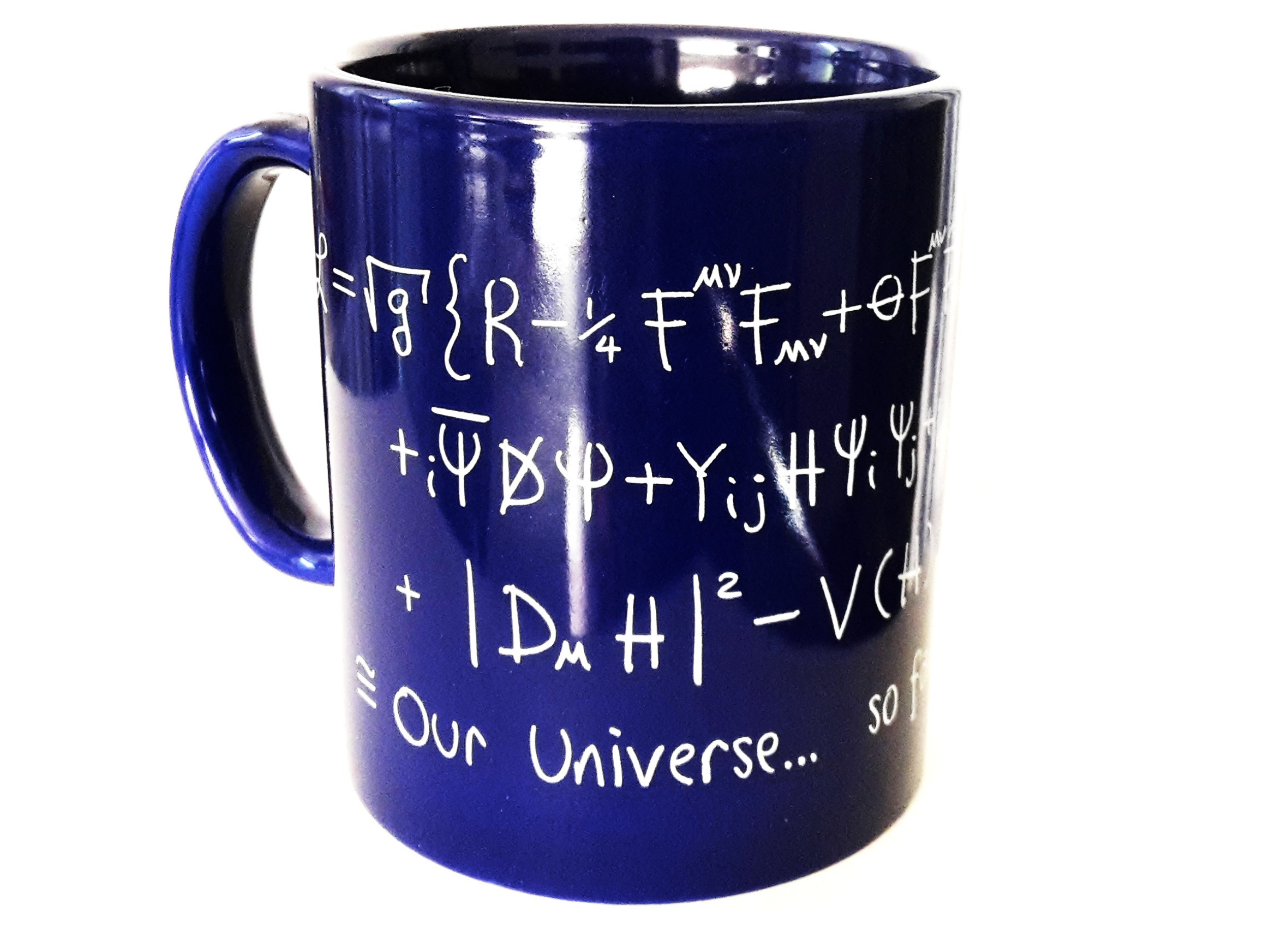}}
\caption{\footnotesize{The standard model on a mug. The first row has the Einstein-Hilbert term for gravity ($\lambda =2$)  and the kinetic and topological terms for the gauge fields ($\lambda =1$) describing the electromagnetic, weak and strong interactions. The second line has the kinetic energy for the matter fields: quarks and leptons $\lambda =1/2$ as well as their (Yukawa) couplings to the Higgs field $H$ ($\lambda =0$). The third line is the kinetic and potential energy for the Higgs field.}}
\label{ra_fig1}
\end{figure}

These are very robust general predictions of relativistic quantum field theories. However, since they are very few, most of our knowledge of field theories is actually based on concrete examples such as QED, QCD,  low-dimensional field theories, etc.

\section{The Standard Model and Beyond}
The standard model is arguably the greatest theoretical development of the past 75 years. It is a very particular QFT based on the gauge theory of the symmetry group $SU(3)\times SU(2)\times U(1)$ with spin $1/2$ matter fields (quarks and leptons) in 3 families fitting in bi-fundamental representations of the factor groups. The theory is described by a simple Lagrangian that can be written on a T-shirt or a mug such as Figure 1.  The experimental evidence is simply spectacular, not only the existence of the relevant particles but the value of the physical observables confirmed with incredible precision over the past 40 years from many experiments. There is no other theory in history that has been tested with such a high precision. Quantities such as the anomalous magnetic moment of the electron fit the experiments with 11 decimal figures and the equivalence principle has also been tested with similar precision with no parallels in any other field of science. The robustness of the standard model is simply impressive. Other features to emphasise about the standard model (SM)  are:
\begin{itemize}

\item{} 
{\it The SM is an EFT}. The non-gravitational part of the Lagrangian is renormalisable and therefore quantum mechanically complete (up to Landau poles). The inclusion of gravity makes it into an effective field theory (EFT) which is well defined  up to scales close to the Planck scale $M_{Planck}=\sqrt{\hbar c/G}\sim 10^{19}$ GeV. The fact that the non-gravitational part of the SM is renormalisable used to be regarded as a positive property. However, it is because of this property that we do not know at which scale the SM ceases to be valid and therefore we have less guidance of what lies beyond the SM. In this sense it could be possible that new physics may only manifest at or close to the Planck scale.

\item{} {\it The SM is simple but not the simplest}. The structure of gauge fields and matter content of the SM is relatively simple (small rank simple gauge groups, matter in bi-fundamental representations). However, there are simpler gauge symmetries (such as just abelian $U(1)$ symmetries or an $SO(3)$ group)  and matter content (only singlets, a single family, etc.) but they do not fit the experiments.

\item{}{\it Rich phase structure}.  The standard model is actually rich enough to illustrate most of the theoretically known phases of gauge theories: Coulomb phase (electromagnetism), Higgs phase (electroweak) and confinement phase (strong interactions). In particular, the three gauge couplings $g_i$ ($i=1,2,3$) measuring the strength of the interactions are such that well controlled perturbative expansions can be defined as long as the quantitiy $\alpha_i=g_i^2/(4\pi)\ll 1$ (for electromagnetism $\alpha_{em}\sim 1/137$  at low energies is the celebrated fine structure constant). This is satisfied for the electroweak interactions at low energies and, thanks to asymptotic freedom, for the strong interactions at higher energies. Extrapolating with the renormalisation group (RG) equations to high energies,  the three gauge couplings tend to approximately unify at energies smaller but  close to the Planck scale and at weak coupling ($\alpha_i<1$). Even though simple extensions of the SM at intermediate scales may modify this behaviour, it is a good hint that an ultraviolet completion of the SM  at weak coupling could exist and therefore a perturbative expansion can hold.  Notice that the couplings $g_i$ themselves are not  too different from one.

\item{} {\it Range of mass hierarchies}. The SM has a wide range of mass scales. From $\Lambda^{1/4}\sim 10^{-3} \rm{eV}\sim 10^{-30} M_{planck}$ relevant for dark energy and approximate range of neutrino masses, to $m_e \sim 10^{-22} M_{planck}$ for the electron mass, to $m_{t,H} \sim 10^{-15} M_{planck}$ for the top and Higgs masses to the Planck scale $M_{planck}$. There is no explanation for this huge hierarchy of mass scales extending for at least 30 orders of magnitude. The only `understood' small mass scale relative to the Planck scale is the QCD scale $\Lambda_{QCD}\sim 10^{-20} M_{planck}$ in which the small factor is determined by dimensional transmutation coming from the logarithmic RG running of the gauge coupling which determines the infrarred cut-off $\Lambda_{QCD}\propto e^{-b/g_3^2}$ with $b$ the QCD beta function coefficient. Here the existence of the large hierarchy is determined by the logarithmic running of the coupling. However,  there is an assumption that the coupling is small but of order one (small changes in $g_3$ generate big changes in $\Lambda_{QCD}$). So far there is no other independent mechanism known  to explain the emergence of a small mass scale and all other hierarchies are left unexplained. 


\item{} {\it The SM is ugly}.  It is important to understand that even though the symmetry principles behind the standard model are very elegant, the concrete experimental realisation that is the standard model itself is ugly in the sense that there are many ($\sim 20$ ) arbitrary parameters, mostly  related to the HIggs couplings, which are not fixed, there exist apparently unnecessary particles expanding  a large range of different mass scales, etc.

\item{} {\it The SM Landscape}. The standard model including gravity imply a landscape of vacua. The Lagrangian of the standard model has a unique solution in four dimensions describing the physics that we know. However, this same Lagrangian allows for an essentially infinite number of solutions in which one of the spatial dimensions is curled into a circle so the space instead of being  the Euclidean $\mathbb{R}^3$ it is $\mathbb{R}^2\times S^1$ with $S^1$ a circle. In \cite{nima} explicit solutions were found fixing the value of the radius of the circle from the parameters of the standard model and using well understood quantum corrections. This provides a concrete realisation of a `landscape' of  a huge number of (2+1 dimensional) universes or multiverse. Notice that usually the existence of a landscape is associated to theories like string theory or higher dimensional gravitational theories that are not yet confirmed by experiment and that the existence of a multiverse is too speculative. However, especially after the discovery of the Higgs, essentially nobody questions the validity of the standard model and yet this experimentally confirmed theory also implies the existence of a landscape of vacua, each vacuum describing a different universe. This makes the idea of the multiverse far less speculative than it is usually presented. 

\item{} {\it The SM is incomplete}. The standard model is almost certainly not complete. It cannot by itself allow for an explanation of dark matter, the density perturbations of the CMB and baryogenesis for instance. Moreover the value of the many parameters of the SM is not understood. In particular the mass of the Higgs is not protected under quantum corrections which tend to bring it to be as high as the limit of validity of the effective field theory, namely $M_{planck}$. The nature of dark energy responsible for the current accelerated expansion of the universe is not understood, especially the fact that it seems to indicate a vacuum energy as small as $\Lambda\sim 10^{-120} M_{planck}^4$. Furthermore,  gravity is described only at the classical or effective field theory level. So the SM is not ultraviolet complete. This is the best evidence we have for the need to go beyond the standard model.

\end{itemize}

In order to search for the new physics that will supersede the SM, we have to explore experimentally all possibilities, increasing the energy, intensity and reach to the highest possible limits. The history of science tells us we are bound to find something. For theorists we can follow several directions:

\begin{enumerate}
\item{} {\it Simplicity}. Add the simplest possible component to the SM (e.g. one extra neutral fermion or boson to be dark matter and/or drive inflation, etc.) and contrast with observations. This is a way to start, at least to eliminate the simplest cases and start building up a more meaningful theory.

\item{} {\it Follow your nose}. Follow aesthetic arguments (usually subjective) as a guideline (e.g. add extra symmetries o dimensions to address dark energy, dark matter or the flavour structure of the SM, consider mechanisms such as the see-saw mechanism to explain smallness of  neutrino masses, etc.).

\item{} {\it Bottom-up}. Use any experimental hint in order to introduce new particles or modifications of the SM that fit data and then use as a guide towards model building (e.g. attempts to explain some astrophysical events from fundamental physics such as a concrete dark matter candidate, attempts to explain some deviations form the SM at colliders data, etc.). 

\item{} {\it Top-down}. Start with a basic theory at high energies and then deduce its implications at lower energies in order to describe the SM and any physics beyond (e.g. Grand Unified Theories (GUT) of the past, extra dimensional theories and supersymmetric theories that address some problems of the SM and/or add the extra possibility of unification of interactions. Since the mid 1980's string theory  has become the prime example of this approach since  on top of addressing those problems it also  includes unification with gravity at the quantum level, providing the explicit ultraviolet complete framework to address physics beyond the SM).

\end{enumerate}

Given the fact that we do not have a clue of what lies beyond the SM, {\it all} of these avenues and potential combinations have to be explored. This is what makes our current state of affairs so interesting. We are facing the problems created by the fact that the SM is so successful experimentally. Once and again the experiments tend to verify its validity  with unprecedented precision and all potential hints of physics beyond the standard model (BSM)  have turned out to be statistical fluctuations, misinterpretation of data or possible to explain within the SM (e.g. faster than light neutrinos, several astrophysical hints for dark matter particles, potential evidence of cosmic strings, primordial gravitational waves, the disappeared $750$ GeV resonance, etc.).

\begin{figure}
\centerline{\includegraphics[height=7cm]{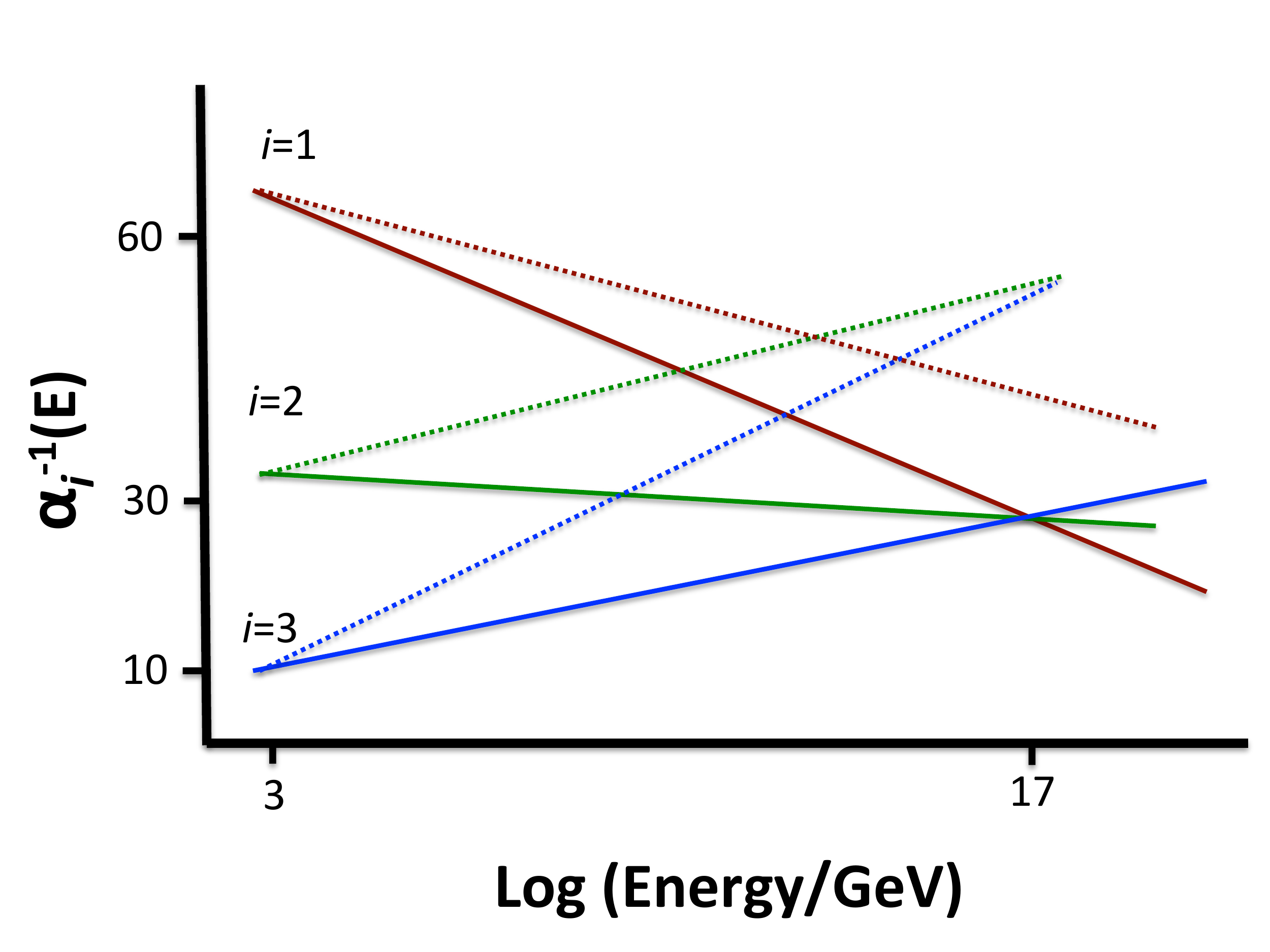}}
\caption{\footnotesize{A sketch of  the running of the SM gauge couplings with energy through the renormalisation group equations. At the TeV scale the three gauge couplings are very different but with increasing energy they tend towards unification. We plot the quantities $\alpha_i^{-1}=\frac{4\pi}{g_i^2}$ against (log of) energy.  Dashed lines correspond to the SM and solid lines to the supersymmetric extension of the SM. Two points are usually emphasised: supersymmetry remarkably unifies the three interactions at an energy close to the Planck scale and even without supersymmetry  the running is towards weaker couplings, raising expectations for a weakly coupled ultra violet completion. Gravity is not included in the plot.}}
\end{figure}

\section{General predictions of String Theory}

Following the case of QFT we may try to extract the general predictions of string theory as known so far (see for instance \cite{witten}).

\begin{itemize}

\item{} {\it One theory, no free parameters.}

Even though over the years 5 consistent supersymmetric string theories were identified, in the mid 1990s it was understood that all of them are different manifestations of one single underlying theory, $M$-theory, that also includes as another weak coupling limit $11$ dimensional supergravity. Uniqueness is an important property that a candidate for a fundamental theory should have. Furthermore, each of the different string theories has one single parameter corresponding to the string length that defines the units and besides that there are no free parameters, another desired property of a fundamental theory. This immediately raises the question of how to determine all the 20 or so free parameters of the standard model taking values in such a huge range. The answer could be dynamical and therefore it is a big challenge to identify the dynamics that selects all the observables we see in nature.

\item{}{\it Extra dimensions.}

Another well known property of consistent string theories is the dimensionality \footnote{Strictly speaking the prediction is on the central charge $c$ of the corresponding conformal field theory (CFT) and the extra contributions to $c$ may not have spacetime interpretation. Also supercritical strings can be considered but proper consideration of the dilaton tadpole needs to be made. It is worth stating that spacetime may not even be the proper arena of the theory  beyond the string scale but its manifestation in terms of 10 or 11 dimensional theories may still be the relevant starting point for exploring low-energy physics with all its hierarchies of scales and  weak couplings.}.  A positive point is that contrary to most theories of physics for which the spacetime dimension is assumed from the start, in string theory it is determined by the condition of criticality of the underlying conformal field theory. The fact that the critical dimension is 10 and not 4 sets a challenge on how to obtain our 4-dimensional world. It immediately indicates that even though the theory may be unique, the number of solutions (giving rise to different universes) most likely cannot be unique. In particular, flat supersymmetric 10-dimensional spacetime should be a solution. Furthermore, the shape and size of the extra-dimensions (known as {\it moduli}) should be determined dynamically. Since they will naturally  affect the physics of the 4-dimensional solution the same 
dynamics that selects them should also  select the physical parameters. Therefore, having more dimensions in a fundamental theory without free parameters could be seen as a positive property that allows the possibility of fixing the many observables of the standard model from the dynamics of the theory. Similar comments apply to the 11-dimensional limit of M-theory. 

\vspace{0.8cm}

\item{} {\it Supersymmetry.}

The third property which is generic in string theories is supersymmetry. Notice that they have either $16$ or $32$ supercharges (supersymmetry generators) which happens to be the maximum number of supersymmetries that allow both just one single graviton and no massless states of helicity larger than 2 (consistent with Weinberg's results mentioned before). Supersymmetry is good for consistency and the absence of tachyons (although non-tachyonic non-supersymmetric solutions are known) but it also raises the challenge of how to find a mechanism to break supersymmetry. Furthermore,  supersymmetric solutions of the theory tend to keep the moduli unfixed. Since moduli represent massless scalar fields in four dimensions, these solutions are very predictive and have been already tested experimentally, namely they  predict long range interactions that should and have not been found and are therefore ruled out by experiment. Furthermore, supersymmetry would imply the existence of supersymmetric partners of the particles we know which have also not been observed. Therefore only non-supersymmetric solutions can be realistic. The scale at which supersymmetry should be broken could be as large as the string scale but also smaller and should be determined dynamically. We can then conclude that determining the value of the moduli and breaking supersymmetry dynamically is a key question for string theory to make contact with the real world.

\item{}{\it Gravity, gauge interactions.}

As  can be seen in the Table, all string theories and 11-dimensional supergravity have in the massless spectrum a symmetric tensor $g_{MN}$ with $M,N=1,2,\cdots 10 (11) $ which, as is well known from field theory, corresponds to  a massless particle of helicity 2 which describing gravitational interactions. This is at the basis of the general statement that string theory predicts gravity in the sense that the existence of a helicity 2 particle is an output rather than an input of the theory. Upon compactification to 4-dimensions the components $g_{\mu\nu},\,\, \mu,\nu =1,\cdots 4$ correspond to the 4-dimensional graviton whereas the $g_{mn},\, \, m,n=1,\cdots 6 (7)$ are scalar fields in 4-dimensions which correspond to some of the  {\it moduli} fields mentioned above.

In the Table it is  seen that gauge fields $A_M^{ij}$ are also present in the heterotic and type I strings. Upon compactification these gauge fields can  give rise to gauge and also matter fields similar to the supersymmetric partners of quarks and leptons. This justifies the fact that before the mid 1990's only these string theories were considered for the search of realistic models. 

\item{}{\it Antisymmetric tensors.}

The new ingredients in the bosonic spectrum of string/M-theory models are the antisymmetric tensors of different indices. They correspond to the only extra fields that give rise to helicities less than 2. In 4-dimensions they are usually not mentioned since they are either equivalent (dual) to scalar fields or they do not correspond to propagating degrees of freedom. But starting from 6-dimensional field theories  these fields correspond to independent propagating fields. They play a role in string theory in at least three ways: 

First, upon compactification, they give rise to pseudoscalar fields or axion-like particles that are the imaginary components of complex moduli fields. This opens up a big window of opportunity for low energy physics since axion-like particles are associated with shifts symmetries of the Peccei-Quinn type that tend to survive at low-energies.

Second, they  indicate the necessary presence of $p$-branes (either D-branes, NS-branes or M-branes). In the same way that the worldline of a particle couples to gauge fields $A_M$ by 
$\int dx^M A_M$ a string couples to a 2-index tensor $\int dx^M\wedge dx^N B_{MN}$ and a $p$-brane to a $p+1$ antisymmetric tensor 
\be
\int dx^{M_1}\wedge dx^{M_2}\wedge\cdots dx^{M_{p+1}} C_{M_1\cdots M_{p+1}}=\int C_{(p+1)}.
\ee
 In the case of Dirichlet or D-branes (coupling to R-R tensors in the Table) they host in their worldvolume open strings with both ends attached to the brane that correspond to 
$U(1)$ gauge fields and, depending on the boundary conditions, a set of $K$ overlapping D-branes can give rise to $U(K), O(K)$ or $Sp(K)$ gauge fields and chiral matter, thereby making the type IIA and IIB string theories viable to have realistic models. Notice in IIA strings there are only even dimensional D-branes whereas in IIB strings there are only odd dimension D-branes (in this case D7 and D3 (or equivalent D9 and D5) branes can host the standard model). Just like the case of charged particles, D-branes, being charged under antisymmetric tensor fields,  have anti branes (which carry the same tension and opposite charge). D-branes and anti D branes preserve different supersymmetries. Therefore the presence of both type of branes in a compactification would break supersymmetry.

The third important role of antisymmetric tensors is that since they are generalisations of gauge fields, they can give rise to non-trivial fluxes. Similar
 to the Dirac magnetic monopole which implies a quantisation of magnetic fluxes over a 2-sphere $\int_{S^2} F_{\mu\nu}dx^\mu dx^\nu=\int_{S^2} B\cdot dS=n$, whenever there is a homological non-trivial surface $\gamma_i$ of dimension $p+2$ there can be fluxes of the field strength  $F_{(p+2)} = dC_{(p+1)}$.
 \be
 \int_{\gamma_i} F_{(p+2)}= n_i
 \ee
 the integers $n_i$ provide an important example of a potentially large set of discrete parameters that can play a role in determining physical observables.

\vspace{5cm}

 \begin{center}
{\bf \large Massless Spectrum of String/M Theories}
\vskip 0.3cm
\begin{tabular}{|c|c|c|c|c|} 
\hline
{\bf Theory} & {\bf Dimensions} &  {\bf Supercharges} & \multicolumn{2}{c|}{\bf Bosonic Spectrum} \\ 
\hline \hline
Heterotic & $10$ & $16$ &  \multicolumn{2}{c|}{$g_{MN},\, B_{MN},\, \phi$}  \\
$E_8\times E_8$ & & & \multicolumn{2}{c|}{$A_M^{ij}$ } \\ \hline \hline
Heterotic & $10$ & $16$ &  \multicolumn{2}{c|}{$g_{MN},\, B_{MN},\, \phi$}  \\
$SO(32)$ & & & \multicolumn{2}{c|}{$A_M^{ij}$ } \\ \hline \hline
Type I & $10$ & $16$ & NS-NS & $g_{MN},\, \phi, A_M^{ij}$  \\ 
$SO(32)$ & & &  R-R & $C_{MN}$ \\ \hline  \hline
Type IIA & $10$ & $32$ & NS-NS & $g_{MN},\, B_{MN},\, \phi$ \\ 
 & & & R-R & $C_M,\, C_{MNP}$ \\ \hline \hline
 Type IIB & $10$ & $32$ & NS-NS & $g_{MN},\, B_{MN},\, \phi$ \\ 
 & & & R-R & $C,\, C_{MN}, \, C_{MNPQ}$ \\ \hline
\hline
11D Supergravity & $11$ & $32$ &  \multicolumn{2}{c|}{$g_{MN},\, C_{MNP}$} \\ \hline
\end{tabular}

\vskip 0.5cm
{\footnotesize{ The five consistent string theories and 11-dimensional supergravity. 
The number of dimensions and supersymmetry generators is given as well as the bosonic  massless content. They all include ($\lambda=\pm 2$ ) gravity ($g_{\mu\nu}$), antisymmetric tensors of different ranks (the $B$ and $C$ fields with $\lambda =\pm1,0$ ) and gauge fields ($\lambda =\pm 1$).}} 
\end{center}

\item{} {\it Infinite tower of massive states.}

The spectrum of string theories also includes an infinite tower of massive states, with masses multiples of the fundamental string scale, and higher spins. All states are related by string creation and annihilation operators.
If the string scale is substantially larger than the TeV scale,  these states would not correspond to any observable particle in the forseeable future. If we are extremely lucky and  the string scale is close to the TeV scale then these states can be the smoking gun for string scenarios to be tested in the near future. 

\item{}{\it No continuous spin representations (CSR).}

As mentioned in Chapter 2, special relativity and quantum mechanics also predict the existence of continuous spin representations which have not been observed. Wigner immediately realised the problem and came up with an (anthropic) argument that if they existed it would require infinite heat capacity to excite them, Weinberg in his textbook simply states (phenomenologically)  that these states have not been observed and then concentrates only on the finite dimensional unitary representations which reduce the Little group to $O(2)$. String theory, at least in its current perturbative formulation, offers a simple explanation for the absence of these states. The reason is very simple. All string states are related by creation and annihilation operators. We can get massless states applying annihilation operators to massive states. Since massive states fit in finite dimensional representations, then massless states should also be in finite dimensional representations and therefore no CSRs \cite{fqt}. Notice that standard field theory does not provide an explanation for this and over the past 75 years different proposals have been made to find physical meaning to CSR's. For  a very recent interesting analysis see for instance \cite{toro}. 

Judging progress in theory by better explanations, this is another success of the theory. If a single evidence for a CSR is found, all classes of string models  studied so far would be ruled out. This reminds the old statement attributed to J.B.S. Haldane  that the theory of evolution could be ruled out if a single fossil of, say a rabbit, is found in the wrong archeological stratum.  CSRs may still be present in a fully-fledged non-perturbative string theory and it may be interesting to investigate their potential implications.



\end{itemize}

\section{Four-dimensional Strings}

As in the case of field theories,  in order to make closer contact to experimental observables, we have to concentrate on more concrete string models with realistic perspectives.

Given the fact that string theories are formulated in 10 (11) dimensions, in order to obtain something like our universe,  solutions (or vacua) have to be found that include the SM at low energies. On this a spacetime of the form:
\be
\mathcal{M}_4\otimes X_6
\ee
should be selected where $\mathcal{M}_4$ is essentially Minkowski space and $X_6$ an internal space. Also the number of supersymmetry generators in 4-dimensions could be from 32,16,8,4,0 (corresponding to what is called $\mathcal{N}=8,\cdots, 1, 0$ supersymmetries respectively). It is well known that only $\mathcal{N}=1,0$ supersymmetries allow for chirality, one of the most important properties of the SM. Therefore this eliminates many possibilities for the extra-dimensional space $X_6$
(such as the simplest cases like toroidal manifolds $X_6=T_6$ ). Solutions with $\mathcal{N}=0$ are physically possible but difficult to handle from our current understanding of string theory. This leaves $\mathcal{N}=1$ as the preferred option. Since we do not observe supersymmetry in nature, this supersymmetry has to be broken, but the breaking scale is not fixed. It may well be that the scale  is as high as the compactification scale and then essentially captures the physics of $\mathcal{N}=0$ compactifications. But it can be as low as our experimental limits of TeV. Most of the work on string compactifications has concentrated on $\mathcal{N}=1$ vacua. 

This process of eliminating solutions which are automatically ruled out by experiment is usually known as ``vacuum cleaning''. Following the original Kaluza-Klein ideas the size of $X_6$ is expected to be as small as to not be detectable experimentally. Progress in the past 10 years has allowed us to actually determine dynamically the shape and size of the extra dimensions and obtain the small sizes expected in the Kaluza-Klein scenario.

It is natural to concentrate on 4-dimensional models (guided by observations and some could also argue anthropic reasons). 
Over the years there has been a big industry of string model building. It is fair to say from the start that at the moment there is not a single fully realistic string model. However, progress has been made in different directions starting from the different known string theories and 11D supergravity.

Clearly the most developed class of models are the heterotic string compactifications on Calabi-Yau (CY) manifolds. At present tens of thousands of models are known with the massless spectrum of the minimal supersymmetric standard model (MSSM) (see for instance \cite{lukas} ). This is non-trivial progress. However, in this theory, given the complicated topology of Calabi-Yau manifolds and the absence of knowledge of explicit metrics, it is very difficult to compute couplings such as physical Yukawa couplings for which not only holomorphic quantities such as superpotentials are needed but also kinetic terms which are non-holomorphic and more difficult to compute. The main problem, however, is a concrete scenario for moduli stabilisation. 

On this type II models are better suited since turning on fluxes goes a long way into stabilising the moduli. Particularly promising are type IIB compactifications for two main reasons: turning on fluxes still keep the compact manifold to be a (conformal) Calabi-Yau and all the knowledge acquired in the past 30 years can be used and second, the fact that there are two different three-index tensors $F_3\sim dC_{(2)}$ and
$H_3\sim dB_{(2)}$ gives rise to a superpotential that fixes most of the moduli. Quasi realistic type II models have been constructed either from intersecting D-branes or D-branes at singularities including three family models containing the SM spectrum and a few extra particles. Only in the past few years fully consistent compact quasi-realistic models have been constructed including moduli stabilisation. But there is much room for finding more realistic models.

There is a more general class of type IIB models known as F-theory in which the dilaton $\phi$ is considered as a modulus for an auxiliary torus and therefore can extend to strong coupling. These models have the interesting property that allow Grand Unified groups such as $E_6, SO(10), SU(5)$ and have potential realistic properties. The mathematics required to fully investigate F-theory models is being uncovered and some progress is expected in a few years but at present there are no concrete models that can be considered realistic and furthermore the computational advantage of weak coupling type IIB models is not maintain and therefore moduli stabilisation is not under control.

An even less developed class of models corresponds to $G_2$ holonomy compactifications in which the standard model has to be at a singularity but not much is known to construct explicit models. However, educated guess has led to interesting developments and concrete scenarios as discussed by Gordy Kane at this meeting.

In summary even though progress has been made and continues to be made there are clearly open challenges.

Rather than concentrating efforts in a search for realistic models, it is better to try to address model-independent issues and extract general scenarios that can be contrasted with experiments.

\subsection{Model Independent Results}

Regarding model independent results, we can mention the following:

\begin{itemize}

\item{} {\it No global symmetries.}

In field theoretical models we can have local and global symmetries. Local symmetries determined the interactions and global symmetries may be used to address some questions such as flavour structure, baryon and lepton number conservation, etc. In string theory there are no global symmetries except for non-compact symmetries such as Poincar\'e invariance or shifts associated to axion fields. The proof is very elegant and based purely on conformal field theory. It essentially states that any potential global symmetry will imply a state in the spectrum with the properties of the gauge field for that symmetry and therefore the symmetry is actually local \cite{banks}. This does not apply to D-branes in flat space and so locally if two sets of D-branes intersect, the local symmetries of one set would be seen as global symmetries of the other. However,  once they are embedded in a proper compactification all symmetries are local and may only remain approximate but not exact global symmetries depending on the structure of the extra dimensions (large volume or large warping). Notice that absence of global symmetries is consistent with standard claims about the absence of global symmetries in theories of gravity coming from the no-hair theorems.

\item{}{\it Small matter representations.}

In string models essentially only fundamental, bifundamental, adjoint, symmetric and antisymmetric elementary representations are realised and (as in the SM) higher dimensional representations are not possible \footnote{Higher level Kac-Moody representations and/or highly Higgsed quiver and particular F-theory models may lead to higher dimensional representations but they are clearly very rare.}. So again if any experiment discovers a representation of dimension, say, 6 of $SU(2)_L$ it would rule out most string constructions known so far.).

\item{}{\it Anomalous $U(1)$s.}

A very stringy low-energy implication is the fact that string theory is consistent and then 
anomaly free  is due to the famous Green-Schwarz mechanism. This mechanism upon compactifications manifests as an extra term in the action that cancels the anomaly of a $U(1)$ gauge group. The net result is that the corresponding $U(1)$ gauge field obtains a mass by eating an axion field in a realisation of the Stuckelberg (but not Higgs) mechanism. The difference is that the symmetry can be broken without a charged field getting a vev. Therefore this provides a low energy effective action mechanism to have approximate global symmetries in string compactifications. Alternatively anomalous and non-anomalous $U(1)$ symmetries may be broken by fields with higher than minimal charge giving rise to discrete symmetries.

\item{}{\it Moduli stabilisation and supersymmetry breaking.}

Independent of the particular constructions,  4-dimensional string models have moduli which need to be stabilised and all start with an original supersymmetric theory and supersymmetry has to be broken. The breaking of supersymmetry may be direct from compactification and therefore the scale of supersymmetry breaking can be as large as the string or Kaluza-Klein scale. In this case it is more difficult to have the low energy effective field theory under control (this is not wrong in itself, it is our own limitation). The more standard construction is to start with $N=1$ supersymmetry compactifications (4 of the 16 or 32 supercharges preserved) using the fact that this is the largest amount of supersymmetry allowed by chirality. In that case the effective filed theory is under better control and supersymmetry may play a role at low energies. Yet the definition of low energy is not specified. These are issues that were open for the first 20 years of string phenomenology with only sporadic attempts and partial success. 

\item{} {\it Moduli and axions.}

Without a concrete mechanism for supersymmetry breaking string models would imply massless moduli which are ruled out by experiment. This means that most string models before 2003 suffer from this problem.
On top of that as it was mentioned before axion-like particles are very generic (coming from antisymmetric tensors) and abound since their number depends on the topology of the extra dimensions, in particular the number of non-trivial homological cycles. Some of the axions become massive with mass of order the gravitino mass but others may remain as essentially massless. 

\item{}{\it Cosmological Moduli Problem.}

In the general class of models in which supersymmetry breaking and moduli stabilisation are addressed (by any unspecified method) there is a very clean prediction: moduli obtain a mass but they survive at low energies and may cause serious cosmological problems \cite{cmp}. Typically their mass is of the order of the gravitino mass (and even smaller) but they would overclose the universe and/or ruin nucleosynthesis if they are lighter than $30$ TeV. That is their mass should be $m>30 $ TeV. This indicates that the supersymmetry breaking scale cannot be that low (which is not bad news given the fact that supersymmetric particles are not yet observed). At first sight it seems  unfortunate that the most generic implication of low energy supersymmetry breaking and moduli stabilisation is actually a problem. On the other hand,  a positive point is that moduli naturally survive at low energies. Even though they couple only gravitationally to the standard model, the fact that they survive at low energies means that there are `stringy' remnants later in the history of the universe  that may have some effects in post-inflationary cosmology.  They change the standard cosmological scenario in which it is assumed that a short period of inflation gives rise to reheating and then the radiation dominated era before big-bang nucleosynthesis. Moduli change this substantially in the sense that after inflation the energy density is dominated by  the moduli oscillations around their minima which behave like matter rather than radiation domination and the true reheating is the late decay of the modulus field. Moduli can contribute and modify standard scenarios of  dark matter, baryogenesis,   dark radiation,  etc.

\item{}{\it Gauge couplings unification.}

Four dimensional strings usually give rise to unification of the gauge couplings even though the gauge group may not necessarily be a simple GUT group. That is, even if the gauge group is a product of simple gauge groups the gauge couplings of non-abelian factors tend to be unified. This is the case in heterotic and models of branes at singularities and most F-theory models (but not necessarily for intersecting brane models) \cite{keith}. Notice that string theories add an extra component to the unification issue since both the unification scale and the gauge coupling itself are quantities that depend on the moduli and have to be determined dynamically. Obtaining precisely the values that fit with the RG running that agrees with the observed gauge couplings at low energies is a very difficult challenge since it has to agree with the independent calculation based on the low-energy spectrum and the scale of supersymmetry breaking (also determined after moduli stabilisation).

\end{itemize}

\subsection{The Landscape}

The landscape and its anthropic implications are the standard source of debate in this field. The existence of energy landscapes is very generic in other areas of science such as in studies of structure and dynamics of atomic and molecular clusters, in protein folding, glasses and supercooled liquids \cite{EnergyLandscapes}. We argued earlier that even the SM has a big landscape. For the string theory landscape I do not have much to add to Joe Polchinski's presentation. What I find important to emphasise here is not the philosophical and anthropic discussions on this issue that have dominated the debate for the past decade and at this conference. Rather I prefer to address the concrete achievements coming out of explicit calculations that address physical questions.

 First, the origin of the landscape was addressing a very important question for string theory models: how to stabilise the moduli \cite{gkp,dk}. The question had been opened for 20 years. Many  of us thought that since the starting point was a supersymmetric compactification, the celebrated non-renormalisation theorems of supersymmetry would imply that only non-perturbative effects could break supersymmetry and most probably we would have to wait for a proper non-perturbative formulation of string theory in order to make progress. The fact that turning on fluxes of antisymmetric tensor fields achieves part of the job is remarkable. These fluxes (in IIB, IIA string compactifications) being quantised give rise to the huge number of solutions.  

Briefly speaking,  CY manifolds have usually hundreds of moduli. They can be classified as the size of 4-cycles (or their dual 2-cycles) which are called K\"ahler moduli labelled by the Hodge numbers
$h_{11}$ and the complex structure moduli corresponding to the size of the non-trivial 3-cycles (and their dual 3-cycles) labelled by the Hodge numbers $h_{21}$. Both $h_{11}$ and $h_{12}$ rank at least in the hundreds  for the known CY manifolds. In IIB strings, since there are two 3-index antisymmetric tensors, their fluxes can thread the corresponding 3-cycles and fix the size of complex structure moduli. For K\"ahler moduli, fluxes do not seem to help but precisely for IIB compactifications there are D7 branes that can wrap these 4-cycles. The size  of these cycles is inversely proportional to the size of the gauge coupling of that theory and non-perturbative effects induce a scalar potential that fixes the moduli. 

Two main scenarios have been developed over the past decade or so differing  on the value of the main quantity in the effective field theory known as the flux superpotential $W_0$ \footnote{Further, less-developed scenarios have been proposed and there may be room for yet further scenarios to emerge in the future, especially in other string theories, in particular non-geometrical fluxes may deserve more studies and fully stringy scenarios of moduli stabilisation may be welcome.}. The original KKLT scenario \cite{kklt}\, works in the regime $W_0\ll 1$ whereas the Large Volume Scenario (LVS) \cite{lvs}\, works for the regime $W_0\sim 1-100$. Both are based on the same ideas but give rise to vey different physics. Even though the moduli are fixed with a negative value of the vacuum energy (cosmological constant) in both cases, the vacuum preserves supersymmetry in KKLT and breaks it in LVS. They both need new ingredients to obtain positive vacuum energy (de Sitter apace). In KKLT the presence of anti D3 branes breaks supersymmetry and adds a positive term to the vacuum energy that can do the uplift. In LVS besides anti D3 branes there are also other options such as T-branes. In both cases the Bousso-Polchinski (BP) \cite{bp} approach to the cosmological constant problem is at work.

Independent of the concrete scenario of moduli stabilisation,  we may inquire about general properties of the landscape. The most concrete prediction of the landscape is that our universe is the outcome of a tunneling event similar to the Coleman-de Luccia bubble nucleation. A direct consequence of this process is that our universe should be open instead of flat or closed unless a long period of inflation occurs. This  has been an open question for many years. Currently,  as we know,  the data tells us that the universe is almost flat and a future precise measure of the curvature of the universe could, in principle,  rule out this scenario. This is a very concrete prediction \cite{susskind}. Achieving the right sensitivity to determine if the universe is open is an important experimental challenge.
This may also have implications  for the low-scale imprints of the CMB.

One general question is if the landscape is finite or infinite. Clearly for classes of vacua with moduli there is a continuum of solutions. But for physically relevant ,vacua it is still not clear if the number of CY manifolds is finite or not although general statements are known (the number of elliptically fibered CY are finite and an overwhelming majority of the known CY manifolds are fibrations). Furthermore general arguments based in studies of sequence of vacua and general mathematical theorems have been given to support  the fact that the number of vacua with `realistic' properties is finite \cite{bobby}.

Another general point is to determine if there are field theories that cannot be included  in the landscape (the `swampland') \cite{swampland}. Even though the landscape of string theories seems to be huge, it is interesting to study cases that will not belong to the landscape in order to extract general results (absence of global symmetries and CSRs could be seen as examples of the existence of the swampland). One of the most popular is the weak gravity conjecture \cite{wgc}. This is a conjecture that gauge-like interactions, including brane-brane interactions are all limited to be stronger than gravity. This is motivated by examples in string theory, the known absence of global symmetries  and black holes physics. This is still only a conjecture but if valid it would not allow many effective field theories, in particular potential candidates to large field inflation, to belong to the landscape.  Also,  some popular proposals for BSM physics to address concrete open questions may be questioned if they could actually have a realisation within string theories or not, for instance models for which the so-called null energy condition is violated seem difficult to be embedded in string theoretical models. This includes proposals for alternatives of inflation and for dark energy.  Moreover,  some of the $f(R)$ scenarios  for modifications of gravity can be contrasted with string theory effective actions, even the original Starobinski proposal for inflation could or could not be incorporated in the landscape, etc. If any of these proposals is found to be in the string  swampland, it does not rule out the proposal but it provides a big challenge to find an ultraviolet completion of the model.

\subsubsection{Supersymmetry Breaking}

Scenarios for moduli stabilisation  can be used to concretely study the breaking of supersymmetry and therefore the spectrum of the supersymmetric particles. See for instance \cite{soft}. The 
 picture that emerges for KKLT and LVS is as follows:

\begin{enumerate}

\item{}
Split spectrum: Mass  $M_{1/2}\geq 1$ TeV  for the fermionic partners of gauge bosons and Higgs (gauginos and Higgsinos)  and $m_0\sim (10-1000) M_{1/2}$ for the scalar partners of quarks and leptons (squarks and sleptons).

\item{}
High scale supersymmetry: All superparticles and gravitino as heavy as $m\sim 10^{11}$ GeV.

\end{enumerate}

The split supersymmetry scenario is realised in KKLT for the SM living on D3 or D7 branes (with a small split $m_0\sim 30 M_{1/2}$) and in the LVS for the SM on D3 branes with a bigger split
($m_0\sim 10^3 M_{1/2}$). High scale supersymmetry is realised in LVS for the SM living on D7 branes. This implies that if, say, light stop particles are detected at LHC it would rule out all these scenarios.  Only particles reachable at LHC  are the gauginos and Higgsinos. There are also limiting cases in which the spectrum is not necessarily split but contracted which can also be tested in the next colliders.

The landscape has positive and not so positive properties for low-energy supersymmetry:

\begin{itemize}
\item{}{\it The Good}

The good news is that the strongest arguments against supersymmetry were always related to the cosmological constant  (CC) problem (SUSY had a chance to solve the CC problem and did not do it and when SUSY is used for the hierarchy problem it neglects the CC problem and any other mechanism that would solve the CC problem may also modify the effects of  SUSY for the hierarchy problem). The landscape just tells us that the Bousso-Polchinski mechanism takes care of the CC for anthropic reasons and then it does not affect any SUSY calculations to address the hierarchy problem in particular soft terms can be trusted. 


\item{}{\it The Bad}

The bad news is that we always hoped that any solution of the CC problem would have further implications at low energies that could be tested (modifications of gravity, etc.) but the landscape solution does not provide any experimental test at low energies.


\item{}{\it The Ugly}

The ugly part is that if anthropic arguments are used for the CC problem then they may also be used for other problems, such as the hierarchy problem, and the need of low-energy SUSY may be essentially irrelevant. But at this stage nature may not care what we consider ugly,  actually the two scenarios mentioned above happen to be  precisely split and high scale supersymmetry which would also rely, at least partially, on the landscape to address the hierarchy problem.

\end{itemize}

\subsubsection{Models of Inflation}

Similarly, a few scenarios have been developed for cosmology, in particular to realise cosmological inflation (see for instance \cite{Baumann}). The inflaton field ranks between brane separation to a K\"ahler or complex structure modulus, axions in different guises, etc.
They all fit well with the most recent constraints from Planck (see the Table). However they disagree on the predictions for the tensor to scalar ratio $r$. If the claim from BICEP 2 ($r\sim 0.2$) would have been correct then all models would have been ruled out except for `axion monodromy inflation'. This implies that these scenarios will be put to test shortly.  Again, this would not be a test of string theory but would test string theory scenarios in a similar way to  the relation of the standard model and its extensions to field theory. This is not as exciting and ambitious as putting to test string theory but it is in the same 
 track as physics that has worked in the past.

The concrete realisation of large field inflation leading to tensor-to-scalar ratio $r$ is very difficult to achieve. String models of inflation have been the main source for candidates, in particular axion monodromy is considered the benchmark model that experimentalists analysed and compared with data regarding values of $r>10^{-3}$. Furthermore a Bayesian analysis (very popular in this conference) of the almost 200 different models of inflation proposed over the past 35 years, selected as the winner a class of string theory models known as K\"ahler inflation. Moreover, it was a cosmologist that had been working in string models of inflation the person that first performed  the detailed study of the foreground for the BICEP2 experiment that lead to the conclusion that BICEP's results were not actually due to tensor modes \cite{flauger}. This study was later confirmed by the detailed analysis of Planck.

Another effect that could be relevant for string models  is dark radiation (relativistic particles such as neutrinos that decouple from thermal equilibrium with standard matter before nucleosynthesis). String models tend to have several hidden sectors and the last modulus to decay after inflation can give rise to observable and hidden sector matter. Relativistic particles in hidden sectors could dominate this decay and then contribute to dark radiation, measure in terms of the effective number of neutrinos $N_{eff}$. We know that $N_{eff}$ is very  much constrained by CMB and nucleosynthesis observations with $\Delta N_{eff}<1$ . On the other hand, if any deviation $\Delta N_{eff}$ from the SM value $N_{eff}=3.04$ is observed in the near future this could be a hint for stringy proposals for dark radiation (such as the axionic partner of the volume modulus).  

\begin{figure}
\centerline{\includegraphics[height=10cm]{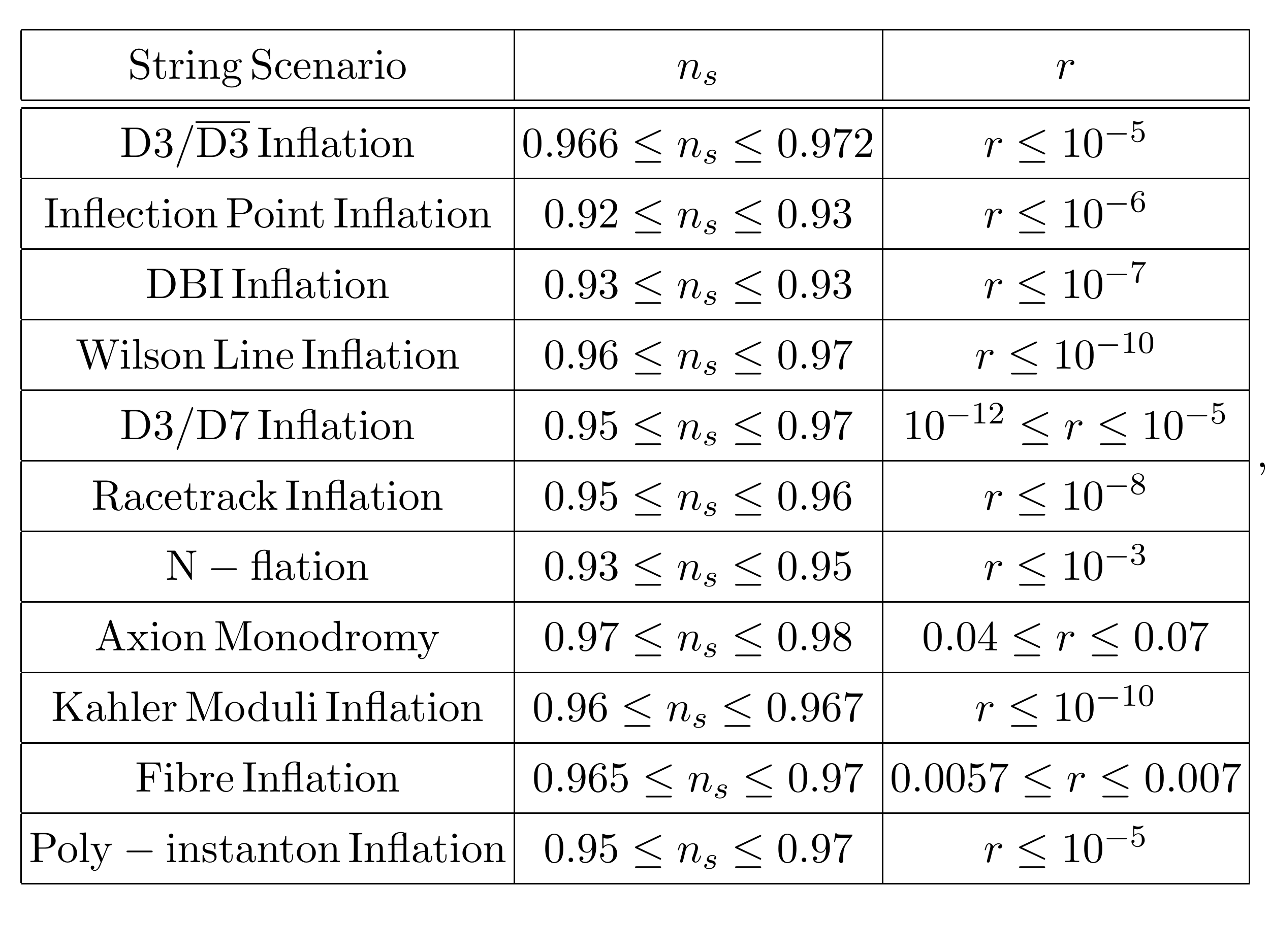}}
\caption{\footnotesize{Table of representatitive models of string inflation (from \cite{bcq}).}}
\label{inflation}
\end{figure}


\section{Criticisms to String Phenomenology}
\begin{itemize} 
\item{}
{\it Not real string theory}.

From a more formal perspective, string theory has many other challenges and it is natural to claim that, before we understand the theory properly, it is better not to address the potential low-energy physical implications, many of them based on effective field theory techniques.
This is a valid argument for some string theorists not to concentrate their research on phenomenological aspects of string theory. However, it is not an argument against other members  of the string theory community to do it. It is well known that the history of scientific discoveries does not follow a logical structure as it happened with the standard model, the model was proposed even before knowing that gauge theories were renormalisable. Nevertheless, 
we may understand the theory better than we  think (at low energies and weak couplings) using all foreseeable  ingredients: geometry, branes, fluxes, perturbative and nonperturbative effects as well as topological and consistency arguments.

String phenomenologists often follow major developments on the formal understanding of the theory and incorporate them into concrete physical models (e.g. CFT techniques for model building and string amplitudes, D-brane/orientifold constructions, F-theory phenomenology, brane/antibrane inflation, physical implications of the open string tachyon, phenomenological and cosmological implications of AdS/CFT, etc.).  
What is however less appreciated is that by  addressing phenomenological questions of string theory, string phenomenologists have actually achieved some of the most important developments in  string theory. First and foremost is mirror symmetry which was essentially identified as a geometrical manifestation of stringy conformal field theories  and later found `experimentally' by just enumeration of Calabi-Yau manifolds relevant for phenomenology \cite{mirror}. Moreover,  the attempt to use mirror symmetry to compute some couplings has led to fundamental discoveries within pure mathematics. Mirror symmetry, after more than 25 years, remains a very active field of pure mathematics. 

Second, $T$- duality, identifying large to small distances in string theory, was discovered by attempts to find interesting cosmological solutions of string theory \cite{tduality}. Also the strong-weak coupling $S$-duality was proposed with the motivation to address the moduli stabilisation problem of string phenomenology \cite{sduality} (actually the widely used terms: $T$-duality and $S$-duality were introduced for the first time in reference \cite{sduality}). Later studies of the implications of these two dualities were the source of the so-called second string revolution of the mid 1990s. This illustrates that the study of string phenomenology has been crucial for the better understanding of the theory itself.

Moreover, some of the most important purely stringy calculations (not just EFT) have been performed in the context of string phenomenology, for instance the string loop corrections to gauge couplings  led to the much studied holomorphic anomaly that has had much impact in formal string and field theory \cite{dkl}.

\begin{figure}
\centerline{\includegraphics[height=10cm]{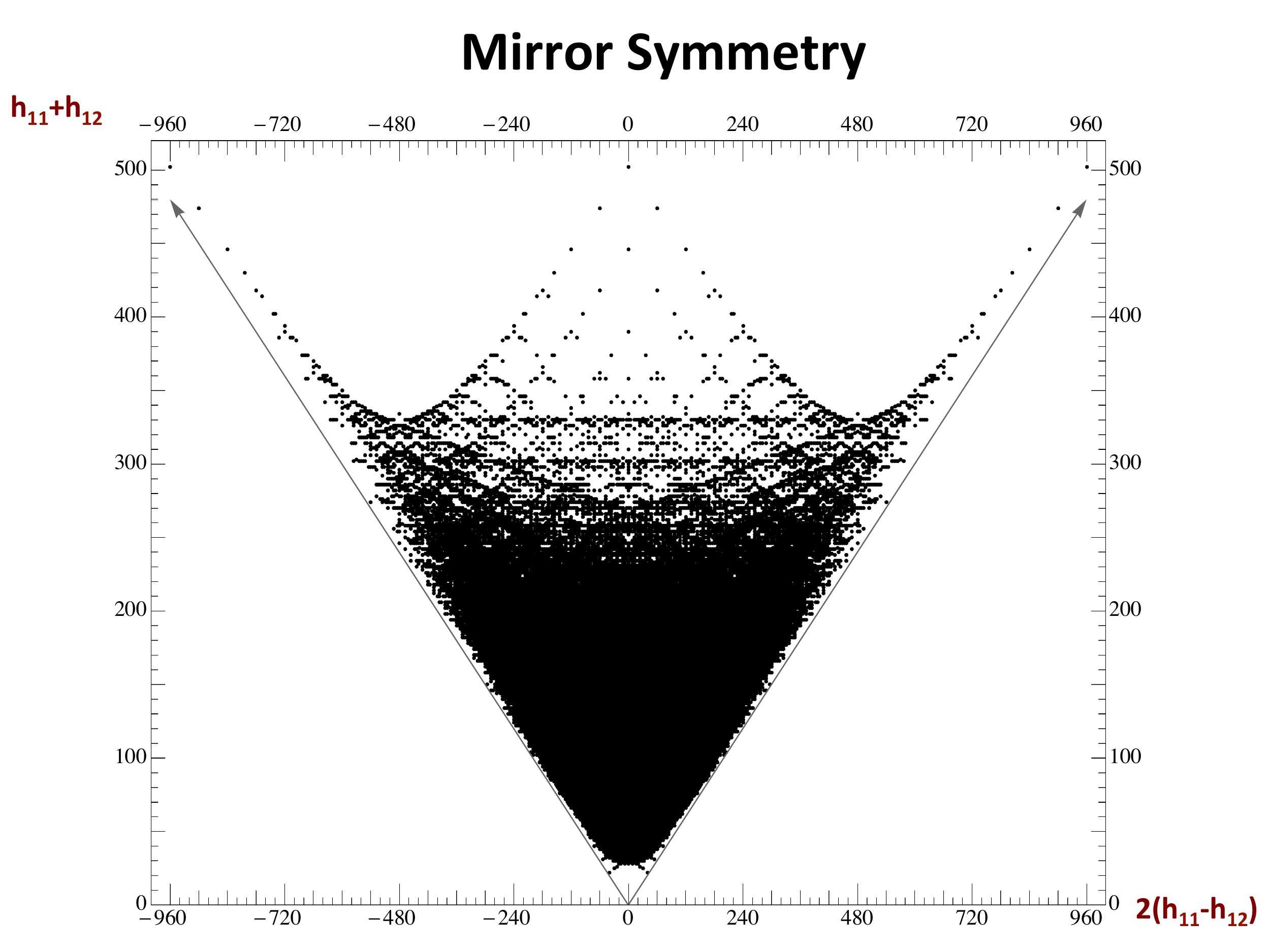}}
\caption{\footnotesize{Evidence for mirror symmetry (the plot is essentially symmetric among interachange of Hodge numbers $h_{11}$ and $h_{12}$. See for instance \cite{cyplot}.}}
\label{ra_fig4}
\end{figure}

\item{}{\it Not real phenomenology}

If there is a sudden experimental discovery or hint of a discovery it will be the theorists working closer to the experimentalists that will make a difference immediately after. String phenomenologists will usually not be the first ones to react (as it has happened  in several occasions) and in that case if a young scientist wants to do phenomenology very close to experiments  then it may be wiser not to enter the complications of string theory and only concentrate on the simplest ways to address the relevant physical questions. This is also a valid argument for part of the community to concentrate purely on phenomenology but with the current stage of affairs it is better to have all different specialisations and tools at hand since we do not know in which direction things will develop. In particular this has been illustrated on the recent advances in the study of the cosmic microwave background and cosmic inflation. Some of the most popular models to compare with experiment are string models (e.g. axion monodromy and DBI inflation).  Furthermore some of the most popular ideas of beyond the standard model (BSM) physics were inspired by string theory (large and warped extra dimensions, split supersymmetry,  axiverse, etc.).

Finally, on a more sociological note, for training young generations that like the challenge of formal and mathematical questions but at the same time  like to keep track of physics developments, this field suits very well. Practitioners enjoy very much this line of research which is a most in the choice of research subjects. Also, even if, in the worst case scenario, string theory is proved wrong or inconsistent in the future, string phenomenologists of all generations will have enough knowledge to optimise versatility and move to other fields. This has actually  been tested over the years by just following the career path taken by some string phenomenologists who either stayed in the field, moved to pure phenomenology, pure cosmology, mostly mathematics, astrophysics, Montecarlo modelling of LHC data and even experimental  high energy physics (besides those who have moved to other fields such as finances or biology). The training they receive is ample enough to be able to adapt to other research areas.

\item{}{\it Not even wrong}

For non-string theorists, the main criticism to the theory itself regards the questions addressed by string phenomenology (not a testable prediction (at least in the short term), failure to detect supersymmetry or other BSM physics hinted by string theory so far, etc.). This criticism is also valid. However, string phenomenologists see these problems as challenges and interesting questions to address in their research which actually have made  the field active and physically interesting over the years. It is actually standard practice in science to fully explore the physical implications of a given  theory and string theory has proven to more than deserve to be explored in this regard despite the fact that it is so difficult to disentangle its physical implications at low energies. In fact one of the key recommendations of Steven Weinberg to young generations \cite{WeinbergNature}\, is precisely to do research in the field that looks not closed and elegant but messy. Given the nature of the problem (to explain all fundamental physical phenomena) we have to be patient and think in the long term. Concrete progress may take several decades or even centuries, given the experimental difficulties, unless we are lucky and several indications materialise that would support the theory or some particular string scenario.

Regarding  time, the recent case of gravitational waves discovery  is an excellent example with a timeframe of 100 years from prediction to experimental confirmation with the techniques for discovering developed some 40 years after the prediction. The 50 years between prediction and discovery of the Higgs particle is another example. Other theories, notably the theory of evolution and the atomic theory, take many decades or centuries before being accepted by the scientific community and even longer to be fully accepted. 

On the other extreme let me mention the case of superconductivity. Historically this was an opposite example in the sense that it was first discovered experimentally (Omnes 1911) and it took almost 50 years for a proper microscopic theory to be developed (Bardeen-Cooper-Schriefer 1957). But it is possible to imagine the situation could have happened in the opposite order. With the knowledge of quantum mechanics and condensed matter systems, theorists could have in principle predicted a superconducting phase of some materials at low temperatures. It did not happen in this way since at that time experiments were easier to perform and it was a difficult challenge to extract macroscopic implications of quantum mechanics from a non-perturbative effect such as the physics of Cooper pairs. Exploring potential physical implications of string theories may be comparable to a potential world in which the study of materials at extremely low temperatures was not available and very talented theorists eventually come up with a prediction of quantum mechanics which implies superconductivity. We can always get lucky.
\end{itemize}

\section{Final Remarks}

 The title of this conference was: Why trust a theory?  The direct answer to this question is clearly that  as scientists we do not trust any theory until it is supported by experimental evidence. But the point of the question, as I understood,  actually relates  to why  theorists (working on string or any other theory) keep working on a theory that has no foreseeable experimental verification. To answer this it is important to separate two issues. One if the theory does not have predictions in principle or if its predictions are not achievable in the short term. In the first case, even though the main goal of science is to have better explanations rather than making predictions, it can be questioned calling it scientific. String theory is clearly not of this type. It actually makes many predictions, especially at or close to the string scale (Kaluza-Klein tower of states, massive string states with specific  behaviour under high energy scattering, etc), which should be the natural case for a theory of quantum gravity. But as we have seen, being less ambitious, some string scenarios have many other potential  implications  at much lower energies which is encouraging.

 In order to eventually be able to rule out the theory,  knowing if  in the UV the theory is weakly  or strongly coupled is important. If the theory is weakly coupled and volumes are large compared to the string scale the perturbative string theory and effective field theory calculations that have been done so far are valid. But if the theory in the UV is strongly coupled we do not understand it well enough to make predictions, except in the regimes open by the strong-weak duality symmetries. Fortunately from what we observe it is not unreasonable to expect the UV theory to be weakly coupled, the naive extrapolation of the gauge couplings from the TeV scale up, even only using the SM particles tends to  point in that direction.  Furthermore the quartic coupling of the Higgs potential is also running towards weak coupling. Even though this may be modified by new physics at scales higher than the TeV scale, it is reasonable to study string theory in the weak coupling regime. Luckily enough this is the regime  in which calculations are  under control. Furthermore having these small parameters (weak couplings and inverse volumes) fits with the general structure of hierarchies we observe in nature.
 
String theory has all the desired ingredients for a fundamental theory of nature (unique, no independent parameters, includes gravity, gauge interactions, matter fields, etc.). It provides a very rich mathematical structure capable to describe our world and deserves to be explored to its limits. There has been continuous progress in 30 years, albeit not as fast as originally expected. There are many open questions  touching issues of gravity, cosmology, phenomenology, astrophysics and mathematics. It is then a fruitful field to do research. We have to be patient and humble and not expect spectacular results that test the theory definitively in a short term. The set of questions to be answered is very large, but exploring different general scenarios and extracting potential  observable implications can give us some guidance.

If in the relatively short term new physics is discovered experimentally it may give indications and guidance for string scenarios but most probably will not give definitive positive test to any particular string scenario. Since string models have to address all BSM questions, searching for correlations among observations may prove useful. It may take several discoveries in different directions in order to get closer to test string scenarios (for instance discovery of low energy supersymmetry and  large tensor modes may eliminate most if not all string scenarios proposed so far, indications of the existence of split supersymmetry and non-thermal dark matter or dark radiation and power loss at large angular scales of the CMB and/or ultra-light dark matter (e.g. $m\sim 10^{-22}$ eV) or keV astrophysical photon excess, axion-photon conversion evidence and/or evidence for very weak gauge interactions would support some concrete scenarios). Also, some general scenarios have very concrete predictions that could be tested (e.g. in LVS the axion partner of the volume modulus is always present and is almost massless providing a model-independent candidate for dark radiation and ultra-light dark matter or dark energy).

Even though predictability is an important aspect to assess a theory it is not its only  {\it raison d'etre}. The main aim of a theory is to provide better explanations of natural phenomena and this is how progress has been  made in science over the centuries. Also   internal consistency is a crucial component for a theory to be seriously considered. As we have seen string theory ranks high on the better explanation aspect. Also, since its beginnings string theory has been subjected to many consistency checks, from anomaly cancellations to many tests of dualities, formal properties of D-branes and connections with mathematics such as K-theory, derived categories, modular forms, mock modular forms, wall crossing, etc. 

Probably some of the most important recent  results have been related with the conjectured AdS/CFT duality, its many checks and potential applications to different  areas of physics, from addressing the information loss paradox, to quark-gluon plasma, cold atoms and quantum criticality in condensed matter systems but the most concrete achievement is  the explicit calculation of the black hole entropy for a class of black hole solutions and their comparison with the famous Bekenstein-Hawking formula $S_{bh}=A/4$ with exact agreement. These developments and many others, including the progress described here in obtaining quasi-realistic string compactifications, give circumstantial evidence and reassures  the researchers working on this field that this is a theory that has deep implications and support the expectations that it has the potential to eventually  provide the ultraviolet completion of the SM. They partially explain why a group of theorists are  committed to do research in this theory despite the fact that  experimental verification will be needed and this may take a long time. 

Let us finish with an optimistic note. Early ideas on black holes go back to the 18th century (Michell, Laplace), they were not seriously considered as physical objects even after Einstein's theory of General Relativity was established and predicted their existence. Following  an almost adiabatic change of perspective during the past 40 years,  supported by several pieces of observational evidence, now their existence is no longer  questioned. Similarly, quarks started life as mathematical entities, later proposed to correspond to physical particles but with the understanding that from their own structure, free quarks cannot be observed in nature, yet physicists were creative enough to identify ways to detect their existence and they are now seen, together with  leptons as the fundamental degrees of freedom for  visible matter.  It is worth recalling Weinberg's statement commenting the early developments on the big-bang model  in his celebrated book  ``The first three minutes"  \cite{weinberg3minutes} that may apply as well to string theory scenarios today:

{\it  This is often the way it is in physics -our mistake is not that we take our theories too seriously, but  that we do not take them seriously enough. It is always hard to  realise that these numbers and equations we play with at our  desks have something to do with the real world.}

\section*{Acknowledgements}
I am grateful to  the organisers, in particular Slava Mukhanov and Richard Dawid for the kind invitation to this interesting meeting. I thank the string phenomenology/cosmology community, especially my collaborators over many years, for shaping my view on the subject. Thanks to Cliff Burgess, Shanta de Alwis, Keith Dienes, Anne Gatti, Luis Ib\'a\~nez,  Dami\'an Mayorga Pe\~na,  Pramod Shukla for comments on the manuscript and to Giovanni Villadoro also for discussions.

\appendix

\section{A timeline of developments in string phenomenology}

In order to illustrate how the field has evolved over the past three decades it may be useful to collect some of the main achievings per decade. This illustrates that the field is evolving maybe slowly but with concrete results and calculations (see also the  String Phenomenology 2014 summary talk by L.E. Ib\'a\~nez at http://stringpheno2014.ictp.it/program.html).

.
\\

\centerline{\bf First Decade: 1985-1994}

\begin{itemize}
\item{} Heterotic string compactifications: (Calabi-Yau) $\times$ (flat 4D spacetime) with $N=1$ supersymmetry.
\item{} $E6$ string phenomenology.
\item{} Simplest dimensional reduction to 4D.
\item{} Gaugino condensation and SUSY breaking. 
\item{} Explicit 4D strings (orbifolds, free fermions, bosonic lattice, Gepner, Kazama-Suzuki models) from exact conformal field theories.
\item{} First quasi realistic string models (CY, orbifolds, free fermions) with three families and including standard model group.
\item{} Evidence for and applications of mirror symmetry.
\item{} Anomalous $U(1)'s$ and field dependent Fayet-Iliopulos terms.
\item{} Computation of threshold corrections to gauge couplings and Gauge coupling unification.
\item{} Introduction of $T$ and $S$ dualities and first attempts at moduli stabilisation.
\item{}Identification of the cosmological moduli problem as a generic low-energy feature of supersymmetric string models.
\item{}First study of soft terms induced from moduli and dilaton supersymmetry breaking.

\end{itemize}

\centerline{\bf Second Decade: 1995-2004}
\begin{itemize}
\item{} Dualities indicating one single string theory.
\item{} D-branes model building: local models (branes at singularities+ intersecting branes).
\item{} Stringy instanton techniques to explicitly compute non-perturbative  superpotentials $W.$
\item{} Large extra dimensions.
\item{} Expanding techniques for model building (heterotic orbifolds, general CYs, G2 holonomy manifolds).
\item{} Fluxes and moduli stabilisation: the landscape (BP, GKP, KKLT).
\item{} Metastable de Sitter (KKLT).
\item{} First computable models of string inflation.
\item{} String inspired alternatives to inflation.
\end{itemize}

\centerline{\bf Third Decade: 2005-2014}

\begin{itemize}
\item{} Large volume scenario (LVS) of moduli stabilisation.
\item{} Instanton calculation tools generalised and expanded (and used for neutrino masses, etc.).
\item{} Local F-theory phenomenology developed reviving string model building with GUT groups (SU(5), SO(10), etc.).
\item{} Big data heterotic models constructions (including many models with the spectrum of the minimal supersymmetric standard model ).
\item{} Discrete symmetries studied in detail and classified as remnants of local gauge symmetries.
\item{} Classes of models of string inflation developed, some including moduli stabilisation and some including possibility of large tensor modes.
\item{} Axiverse (axion like particles covering a large spectrum of masses and couplings)  and cosmic axion background introduced.
\item{} Global model building developed including first compact CY compactifications with moduli stabilisation (including de Sitter vacua) 
\item{} Concrete studies of global F-theory models (including SM-like models, extra $U(1)$'s, etc.).
\item{} Concrete calculations regarding quantum (loop and $\alpha'$) corrections to K\"ahler potential and their phenomenological and cosmological implications.
\item{} Realisations of de Sitter vacua from purely supersymmetric effective field theories (including also constrained superfields).
\item{} Exploring general constraints to large field inflation, including weak gravity conjecture.
\end{itemize}

\section{Ten open questions for BSM }

Any attempt towards physics beyond the standard model (BSM)  has to address at least one of the  questions below. What makes string models so challenging is that if it fails in {\it one} of these issues then the model is ruled out. 
Even though there are string models that address several of these questions a fully realistic string model does not exist yet. Non-stringy models usually concentrate on a small subset of the questions and in particular question 1 is often ignored. 
 
\begin{enumerate}
\item{} Ultraviolet completion
\item{} Gauge and matter structure of the SM (origin of basic degrees of freedom and corresponding quantum numbers).
\item{} Origin of values of SM parameters,  including  hierarchy of scales, masses (including neutrino masses), quark mixing (CKM matrix), lepton mixing (PMNS) and right amount of CP violation avoiding unobserved flavour changing neutral currents.
\item{} Strong CP problem.
\item{}Hierarchy of gauge couplings at low energies potentially unified at high energies
\item{} Almost stable proton with a quantitative account of  baryogenesis (including appropriate CP violation).
\item{} Source of density perturbations in the CMB (realisations of inflation or alternative early universe cosmologies)
\item{} Explanation of dark matter (avoiding over-closing).
\item{} Absence or small amount of dark radiation consistent with observations ($4\geq N_{eff}\geq 3.04 $)
\item{} Explanation of dark energy (with equation of state $w=p/\rho \sim -1$).

\end{enumerate}

\section{Some open challenges for string phenomenology}

\begin{enumerate}

\item{} Computation of (holomorphic) Yukawa couplings and  quantum (string and $\alpha'$ ) corrections to K\"ahler potentials for moduli and matter fields for different string theories and F-theory.
\item{} Moduli stabilisation for heterotic and F-theory models.
\item{} Stringy control of moduli stabilisation (including both open and closed string moduli) and de Sitter uplift.
\item{} Explicit model of inflation with calculational control, especially for the case with large tensor modes.
\item{}Better control on non-supersymmetric compactifications as well as supersymmetry breaking terms.
\item{} Explicit constructions of $G_2$ holonomy chiral models from 11-dimensions.
\item{} Model independent information of flavour issues including neutrino masses.
\item{} Stringy mechanisms of baryogenesis.
\item{} Model independent studies of stringy dark matter candidates (from visible and hidden sectors).
\item{} Explicit construction of a fully realistic string model or scenario (including early universe cosmology, dark matter candidates, mechanism for baryogenesis, etc.) that reduces to the SM at low energies.
\end{enumerate}

\end{document}